\documentclass[%
 reprint,
 amsmath,amssymb,
 aps,
]{revtex4-2}

\usepackage{graphicx}
\usepackage{dcolumn}
\usepackage{bm}
\usepackage{kotex}
\usepackage{braket}
\usepackage{multirow}
\usepackage{color}

\begin{document}

\preprint{APS/123-QED}

\title{Impact of High-Brightness Entangled Photon Pairs on CHSH Inequality Experiment}

\author{Jin-Woo Kim}
 \altaffiliation[Also at ]{The School of Electrical Engineering, Korea Advanced Institute of Science and Technology, Daejeon, Republic of Korea}
\author{Suseong Lim}%
\affiliation{%
 The School of Electrical Engineering, Korea Advanced Institute of Science and Technology, Daejeon, Republic of Korea
}%
\author{Heonoh Kim}%
\affiliation{%
 Satellite Technology Research Center, Korea Advanced Institute of Science and Technology, Daejeon, Republic of Korea
}%
\author{June Koo Kevin Rhee}%
 \email{rhee.jk@kaist.edu}
\affiliation{%
 The School of Electrical Engineering, Korea Advanced Institute of Science and Technology, Daejeon, Republic of Korea
}%

\date{\today}

\begin{abstract}
Verifying the violation of Bell's inequality is one of the most representative methods to demonstrate that entangled photon pairs prepared in a quantum optics-based system exhibit quantum properties. While experiments on Bell inequality violations have been theoretically well-established and extensively conducted to implement various quantum information technologies in laboratory settings, mathematical modeling for accurately predicting the distribution of high-intensity entangled photon pairs in high-loss environments remains an issue that requires further research. As the brightness of the entangled photon pairs increases, the influence of multi-photon effects becomes more significant, leading to a decrease in the CHSH value $S$ and also a reduction in the standard deviation of the CHSH value $\Delta S$. Therefore, a new analysis of the $(S-2)/\Delta S$ value is required to more precisely confirm the degree of CHSH inequality violation including the reliability of $S$. In this paper, we propose a mathematical model to predict the $(S-2)/\Delta S$ value as a function of the brightness of the entangled photon pair source, and we also suggest the need to optimize the brightness of this source. Additionally, we provide experimental evidence supporting this model. The experiment confirms that when the mean photon number is $\mu=0.026$ in an entanglement distribution setup with a total loss of $-19.03$ dB, the CHSH value drops to 2.69, while the $(S-2)/\Delta S$ value increases to 60.95.
\end{abstract}

\maketitle

\onecolumngrid

\section{\label{sec:level1}Introduction}
Nonlocality indicates to the quantum property that allows for physical interactions to influence each other even when spatially separated. The theoretical inquiry of such nonlocality has been intertwined with the history of quantum mechanics since its inception\cite{einstein1935can, bell1964einstein}. On the other hand, experimental verification of nonlocality has significantly contributed to our understanding of  nature\cite{freedman1972experimental,aspect1982experimental,pan2000experimental}. Entangled photon pairs are a clear example of these quantum properties, allowing for maximum violation of the CHSH inequality. Entangled photon pairs are not only interesting from the perspectives of quantum mechanics and quantum information theory, but they are also considered a key element in various engineering applications, such as quantum key distribution (QKD), quantum teleportation, quantum imaging, and quantum networks\cite{bennett1992quantum, bennett1993teleporting, pittman1995optical, yu2020entanglement}. Consequently, researchers have been interested in verifying how entangled the prepared photon pairs are in actual experiments, and creating higher-quality entangled photon pairs has become an important research issue. In the past, detectors for single-photon detection, such as photo multiplier tubes, exhibited very low detection efficiency.  However, recent advancements have led to the use of detectors like superconducting nanowire single-photon detectors, which show low error rates and efficiencies exceeding 90$\%$. These developments have helped close loopholes in the proof of nonlocality through various experiments\cite{pearle1970hidden, giustina2015significant}, including long-distance Bell inequality violation tests. Furthermore, a comprehensive understanding of single-photon-level entangled photon pairs has also progressed significantly.

As research is conducted to explore the physically essential issues related to entanglement, studies aimed at enabling the practical application of this entanglement as a technology have steadily advanced. Beginning with the first 13km free-space BBM92 experiment in 2005, long-distance QKD was conducted over a distance of 144 km in 2007, followed by a satellite QKD between a satellite and a ground station at a distance of 1200 km in 2017\cite{peng2005experimental, ursin2007entanglement, yin2017satellite}. This marked the beginning of efforts to establish a quantum network among the satellites and the ground stations, shifting from theoretical proposals to practical implementations. Currently, long-distance applications of quantum information technology are being carried out on Earth, but in the future, technologies in space are also being considered and planned\cite{cao2018bell,bedington2017progress}. However, the progress of these technologies is not without its challenges. The transmission of entangled photon pairs through optical fiber results in a loss of 0.2 dB per km, making long-distance transmission inherently difficult. While it may be anticipated that free-space transmission of photons would only suffer from atmospheric attenuation or that transmission from space would experience negligible attenuation, in practice, the transmission is primarily constrained by diffraction effects associated with the specifications of telescope size\cite{aspelmeyer2003long}. Furthermore, as the distance increases, the transmitted beam tends to fluctuate, requiring tracking and correction of the beam position. Taking these issues into account and constructing the satellite-to-ground QKD system, the attenuation rates in free space at distances ranging from 500 km to 1200 km are reported to be between 20 dB and 50dB\cite{liao2017satellite, ren2017ground, yin2020entanglement}. Such high attenuation rates constrain adequate transmission at the single-photon level, as maintaining single-photon intensity complicates the transmission process. Conversely, increasing the intensity to mitigate attenuation compromises the quality of transmission. These concerns highlight the necessity for proper analyses and modeling of the transmission of high-brightness entangled photon pairs.

The mathematical modeling of high-brightness entangled photon pairs began in the 1990s with the experimental realization of such pairs and has since become a consistent theme of research \cite{kwiat1995new}. These high-brightness entangled photon pairs no longer manifest as single photons but rather generate multi-photon effects \cite{simon2003theory,eisenberg2004quantum}. Unlike weak coherent laser light produced by attenuated lasers, entangled photon pairs follow photon number statistics that reflect a two-mode squeezed vacuum state rather than a coherent state\cite{navarrete2012enhancing,roy2018response}. Moreover, the resulting photons do not all exhibit the same polarization state; while the pairs maintain correlation, they exhibit random polarization states. This characteristic ultimately grants entangled photon pairs a unique high-dimensional quantum state defined by superposition. Understandably, this research has expanded in various directions related to the distribution of entangled photon pairs. However, analyzing entanglement distribution under general circumstances has proven to be computationally challenging, leading to studies conducted under very extreme conditions \cite{durkin2004resilience,caminati2006nonseparable,takesue2010effects,takeoka2015full}, analyses of limited application scenarios \cite{ma2007quantum,semenov2010entanglement}, or numerical computations which accept a degree of error \cite{yoshizawa2012evaluation}. Recently, several mathematical models addressing entanglement distribution in a general regime have been proposed to fill these gaps. However, it is unfortunate that the significance and existence of such modeling have not been adequately acknowledged within the field \cite{brewster2021quantum,kim2024strategy}. In experimental settings, the development of quantum sources and the analysis of their characteristics are driven by the requirement for ultra-bright entangled photon pair source to enhance the performance of practical applications \cite{steinlechner2014efficient,yin2017EPP,cao2018bell,ecker2021strategies,park2024ultrabright}.  However, the focus has primarily been on increasing the generation rate of entangled photon pairs and evaluating the visibility achieved in each research group's experimental setup. It is well established in previous theoretical studies that there is an inherent trade-off between the quantity and quality of entangled photon pairs. Therefore, the primary aim lies in the development of mathematical modeling that can predict the optimized intensity of the entangled photon pair source required for the intended applications or systems, alongside establishing empirical evidence for such modeling. It is anticipated that the mathematical modeling of high-brightness entangled photon pair distribution will have significant applications in the optimization of quantum key distribution (QKD) and the development of quantum networks.

In this study, we introduce the importance of considering not only the value of CHSH inequality violation represented by $S$, but also the standard deviation $\Delta S$ and its physical significance \cite{weihs1999experiment}. The inclusion of the standard deviation is crucial, as it is ultimately influenced by the visibility and brightness of the experiment. While this aspect has been analyzed previously, as brightness increases, the contribution of multi-photon effects becomes more pronounced. Consequently, the statistical properties of the measured values can no longer be treated as those at the single-photon level, necessitating more precise mathematical modeling. For this reason, this paper presents a mathematical model that predicts the results of CHSH experiments using our previous research outcomes related to measurement statistics of entangled photon pair distribution considering the multi-photon effects. Additionally, we propose the necessity of optimizing the intensity of the entangled photon pair source. The experiments used a Type II periodically poled potassium titanyl phosphate (PPKTP) to construct a Sagnac interferometer for generating entangled photon pairs. It involves varying the brightness of the entangled photon pairs in environments with different attenuation rates. To minimize other factors influencing the visibility of the entangled photon pairs, excluding brightness, the experiments were conducted with a pump beam of 1.13 mW, resulting in a visibility of approximately 0.985. The angle of the measurement basis was rotated to verify whether the coincidence values were balanced, indicating the establishment of an entangled photon pair source. Subsequently, the results from the CHSH inequality violation experiments confirmed that the proposed mathematical modeling adequately accounts for the experimental findings. 

\section{Results}
\subsection{Theoritical modeling}
The violation of Bell inequalities are used not only to demonstrate the nonlocal properties of distributed photon pairs but also to evaluate how closely entangled photon pairs encoded as bipartite qubits approach an ideal Bell state\cite{bell1964einstein, freedman1972experimental, aspect1982experimental}. While the methods for conducting Bell inequality experiments vary slightly depending on the objectives of the research, this paper employs the CHSH inequality, which is commonly used to verify polarization-based entanglement\cite{clauser1969proposed}. In the experimental setup, the modes of the distributed entangled photon pairs are designated as Alice mode (A) and Bob mode (B). It is assumed that both A and B can simultaneously measure detectors aligned to perpendicular polarization. Under this experimental configuration, the CHSH value for the distributed $\psi^-$ or $\phi^+$ Pauli states follows
\begin{subequations}\label{eq:CHSH}
\begin{eqnarray}\label{eq:chsh}
    S=|E_{1,1}-E_{1,2}+E_{2,1}+E_{2,2}|,
\end{eqnarray}
\begin{equation}\label{eq:dchsh}
    \Delta S=\sqrt{[\Delta E_{1,1}]^2-[\Delta E_{1,2}]^2+[\Delta E_{2,1}]^2+[\Delta E_{2,2}]^2}.
\end{equation}
\end{subequations}
Here the $S$ is described with the correlation parameters as
\begin{equation}\label{eq:estimate}
    E_{i,j}=\frac{N_{\parallel}(\varphi_{Ai}-\varphi_{Bj})-N_{\perp}(\varphi_{Ai}-\varphi_{Bj})}{N_{\parallel}(\varphi_{Ai}-\varphi_{Bj})+N_{\perp}(\varphi_{Ai}-\varphi_{Bj})},
\end{equation}
where A and B prepare their measurement bases at angles of $\varphi_{A1}=0^\circ$, $\varphi_{A2}=45^\circ$, $\varphi_{B1}=22.5^\circ$, and $\varphi_{B2}=67.5^\circ$ respectively. The subscripts of $E_{i,j}$ correspond to the indices indicating the measurement bases used by A and B. The $N_{\parallel}=N_{A_+,B_+}+N_{A_-,B_-}$ represents the sum of the coincidence counts from the detectors $A_+,B_+$ and $A_-,B_-$. In contrast, $N_{\perp}=N_{A_+,B_-}+N_{A_-,B_+}$ signifies the sum of the coincidence counts from the detectors $A_+,B_-$ and $A_-,B_+$. We note that discarding N-fold coincidences in Bell inequality violation experiments reflects the intent of the experimenter, potentially leading to loopholes. Therefore, the appropriate treatment of coincidence counts is represented as a linear combination of individual N-fold coincidence counts $C_{A,B}$, borrowing from the Squash model, such that $N_{A_+,B_+}=C_{A_+,B_+}+\frac{1}{2}C_{A_+,B_\pm}+ \frac{1}{2}C_{A_\pm,B_+} + \frac{1}{4}C_{A_\pm,B_\pm}$ \cite{beaudry2008squashing,tsurumaru2008security,semenov2010entanglement,kravtsov2023security}.

Generally, fundamental quantum information experiments require very low generation rates of entangled photon pair sources, typically at the single-photon level. In such cases, the CHSH value is proportional to the visibility of the coincidence fringe. Factors influencing the visibility in CHSH experiments include the quality of the entangled photon source, the performance of individual components within the system configuration, the alignment of detection efficiencies among detectors, and noise characteristics, such as dark counts and afterpulse. When the statistical error of the measured values follows a Poisson distribution, the CHSH value can be expressed as $S=4E(0)$, where $E(0)=|\frac{N_{\parallel}(0)-N_{\perp}(0)}{N_{\parallel}(0)+N_{\perp}(0)}|$, and the standard deviation is given by $\Delta S=\sqrt{\frac{2-2E^2(0)}{N_{\parallel}(0)+N_{\perp}(0)}}$ \cite{weihs1999experiment}. A higher standard deviation of the CHSH value, $\Delta S$, indicates lower accuracy, suggesting that larger coincidence counts are necessary to reduce the standard deviation. Notably, achieving a higher maximum coincidence count requires increasing the brightness of the entangled photon pairs, which ultimately necessitates consideration of the multi-photon effect and can lead to a reduction in visibility. Consequently, a trade-off exists between the brightness of the source and the accuracy of the CHSH value. For this reason, when conducting experiments to violate Bell inequalities, it is essential to consider both the CHSH value and its accuracy together. Unfortunately, in real experiments, the statistical errors of correlation measurements do not always follow Poisson statistics, especially when the brightness of the photon pairs increases. This implies that in applications within high-loss environments, there is a need to analyze the measurement statistics of correlations at high brightness. The investigation focuses on the impact of the multi-photon effect of high-brightness entangled photon pair sources on the quantity $\frac{S-2}{\Delta S}$.

The photon count statistics of correlations considering the multi-photon effect can be derived from the Hamiltonian of spontaneous parametric down-conversion (SPDC). To create entangled photon pair sources, a Sagnac interferometer is commonly used, and the Hamiltonian for generating the Bell state $\psi^-$ form of entangled photon pairs is given by\cite{simon2003theory,eisenberg2004quantum}
\begin{equation}\label{eq:hamiltonian}
    \hat{H}=i\kappa (\hat{a}^{\dagger}_{A_H}\hat{a}^{\dagger}_{B_V}-\hat{a}^{\dagger}_{A_V}\hat{a}^{\dagger}_{B_H})+h.c.
\end{equation}
where the subscripts of the creation operators indicate the horizontal (H) and vertical (V) polarization state modes for Alice and Bob. $\kappa$ represents the nonlinearity constant, which affects the generation rate of the entangled photon pairs. By calculating the time evolution using the given Hamiltonian as $U=e^{-i\hat{H}t/\hbar}$, the quantum state considering the multi-photon effect is 
\begin{equation}\label{eq:psi}
    \ket{\Psi}=\frac{1}{\cosh^2\gamma}\sum_{n=0}^{\infty}\frac{\tanh^2\gamma}{n!}\left(
    \hat{a}^{\dagger}_{A_H}\hat{a}^{\dagger}_{B_V}-\hat{a}^{\dagger}_{A_V}\hat{a}^{\dagger}_{B_H}
    \right)^n\ket{0},
\end{equation}
where $\gamma=\kappa T_{int}$ is the nonlinear gain during the interaction time $T_{int}\leq t_c$ that is shorter than the coherence time of the laser $t_c$. Here, the average photon number is given by $\mu=\sinh^2\gamma$. The $\gamma$ is influenced by the properties of the nonlinear crystal, phase matching conditions, and the characteristics of the laser, as well as the laser power. Unlike other conditions determined by applications and system design, $\gamma$ is proportional to the intensity of the laser, which is proportional to the square root of the laser power $\sqrt{P}$.

The distributed quantum state, as shown in Eq. \ref{eq:psi}, experiences attenuation due to the channel. The attenuation rate varies depending on whether the distribution channel uses optical fibers or transmits in free space, the attenuation of the distributed quantum state can fundamentally be described using the beam splitter (BS) model \cite{caminati2006nonseparable}. The distributed quantum state yields single counts and coincidence counts based on the choice of measurement bases by the detectors. In this study, the analysis and experiments were conducted using a threshold detector, a practical detector, instead of a number-resolving detector. The processing method for the measured information may involve handling double click events, which can be approached through the squash model or the discard model; however, for Bell inequality violation experiments, the squash model must be employed, and this was taken into consideration during the analysis and experiments \cite{semenov2011fake}. In the context of entanglement distribution taking the multi-photon effect into account, there are a total of 16 measurement outcomes. These measurement probabilities can be expressed as a function of the angular difference $\theta$ between Alice and Bob's measurement bases, and they can be represented as a linear combination of a newly defined $Q_j$-function\cite{kim2024strategy},
\begin{equation}\label{eq:Q-func}
    Q_j(\theta)=\frac{ABCD}{ABCD+G^2-G(AD+BC)\cos^2\theta-G(AC+BD)\sin^2\theta},
\end{equation}
where the constants $A,B,C,D$ are given as values related to the attenuation rates $\tau_1$ and $\tau_2$ according to the indexing of the $Q_j$-function, as shown in Table. \ref{tab:app_P}.

\begin{table}[t]
\caption{Each $Q_j$-functions are given with Eq. \ref{eq:Q-func} when there coefficients $A,B,C,D$ are given as follows.} 
\renewcommand{\arraystretch}{1} 
\resizebox{0.5\columnwidth}{!}{
\begin{tabular}{c cc cc c c cc cc}
\hline\hline
\multirow{2}{*}{\textbf{$Q_j$}} & \multicolumn{2}{c}{\textbf{Alice}}          & \multicolumn{2}{c}{\textbf{Bob}}            &  & \multirow{2}{*}
{\textbf{$Q_j$}} & \multicolumn{2}{c}{\textbf{Alice}}          & \multicolumn{2}{c}{\textbf{Bob}}            \\ \cline{2-5} \cline{8-11} 
                                      & \multicolumn{1}{c}{\textbf{A}} & \textbf{B} & \multicolumn{1}{c}{\textbf{C}} & \textbf{D} &  &                                       & \multicolumn{1}{c}{\textbf{A}} & \textbf{B} & \multicolumn{1}{c}{\textbf{C}} & \textbf{D} \\ \cline{1-5} \cline{7-11} 
\textbf{$Q_1$}                           
& \multicolumn{1}{c}{$1$} & $1$  & \multicolumn{1}{c}{$1$} & $1$          &  & \textbf{$Q_9$}                           
& \multicolumn{1}{c}{1} & $1-\tau_1$ & \multicolumn{1}{c}{$1$} & $1-\tau_2$ \\[1.2ex] \cline{1-5} \cline{7-11} 
\textbf{$Q_2$}                           & \multicolumn{1}{c}{$1-\tau_1$}          & $1$         & \multicolumn{1}{c}{$1$}          & $1$          &  & \textbf{$Q_{10}$}                           & \multicolumn{1}{c}{$1-\tau_1$}          & $1-\tau_1$          & \multicolumn{1}{c}{$1$}          & $1$          \\[1.2ex] \cline{1-5} \cline{7-11} 
\textbf{$Q_3$}                           & \multicolumn{1}{c}{$1$}          & $1-\tau_1$          & \multicolumn{1}{c}{$1$}          & $1$          &  & \textbf{$Q_{11}$}                          & \multicolumn{1}{c}{$1$}          & $1$          & \multicolumn{1}{c}{$1-\tau_2$}          & $1-\tau_2$          \\[1.2ex] \cline{1-5} \cline{7-11} 
\textbf{$Q_4$}                           & \multicolumn{1}{c}{$1$}          & $1$          & \multicolumn{1}{c}{$1-\tau_2$}          & $1$          &  & \textbf{$Q_{12}$}                          & \multicolumn{1}{c}{$1-\tau_1$}          & $1-\tau_1$          & \multicolumn{1}{c}{$1-\tau_2$}          & $1$          \\[1.2ex] \cline{1-5} \cline{7-11} 
\textbf{$Q_5$}                           & \multicolumn{1}{c}{$1$}          & $1$          & \multicolumn{1}{c}{$1$}          & $1-\tau_2$          &  & \textbf{$Q_{13}$}                          & \multicolumn{1}{c}{$1-\tau_1$}          & $1-\tau_1$          & \multicolumn{1}{c}{$1$}          & $1-\tau_2$          \\[1.2ex] \cline{1-5} \cline{7-11} 
\textbf{$Q_6$}                           & \multicolumn{1}{c}{$1-\tau_1$}          & $1$          & \multicolumn{1}{c}{$1-\tau_2$}          & $1$         &  & \textbf{$Q_{14}$}                          & \multicolumn{1}{c}{$1-\tau_1$}          & $1$          & \multicolumn{1}{c}{$1-\tau_2$}          & $1-\tau_2$          \\[1.2ex] \cline{1-5} \cline{7-11} 
\textbf{$Q_7$}                           & \multicolumn{1}{c}{$1-\tau_1$}          & $1$          & \multicolumn{1}{c}{$1$}          & $1-\tau_2$          &  & \textbf{$Q_{15}$}                          & \multicolumn{1}{c}{$1$}          & $1-\tau_1$          & \multicolumn{1}{c}{$1-\tau_2$}          & $1-\tau_2$          \\[1.2ex] \cline{1-5} \cline{7-11} 
\textbf{$Q_8$}                           & \multicolumn{1}{c}{$1$}          & $1-\tau_1$          & \multicolumn{1}{c}{$1-\tau_2$}          & $1$          &  & \textbf{$Q_{16}$}                          & \multicolumn{1}{c}{$1-\tau_1$}          & $1-\tau_1$          & \multicolumn{1}{c}{$1-\tau_2$}          & $1-\tau_2$         \\[1.2ex]
\hline\hline
\end{tabular}
}\label{tab:app_P}
\end{table}

\begin{figure}
    \centering
    \includegraphics[width=\linewidth]{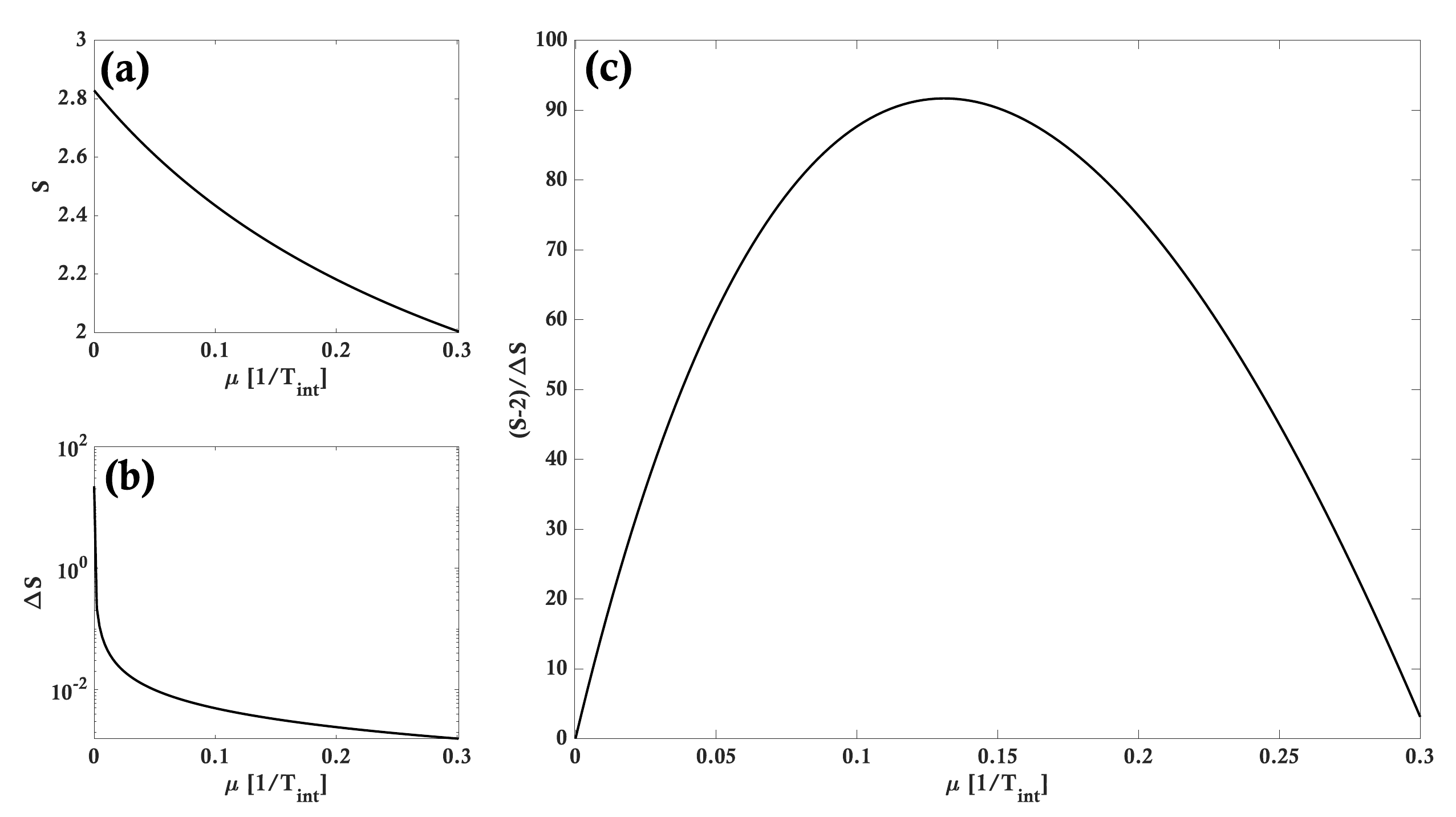}
    \caption{The predicted results of the Bell inequality violation experiment, assuming the presence of entangled photon pairs with a mean photon number $\mu$ over a single temporal mode. External environmental factors that could influence the CHSH experiment, such as dark counts, afterpulse, and stray light, are excluded from consideration to focus on the effects of the multi-photon effect. To match with experimental values discussed later, the channel environment during the distribution of the entangled photon pairs is configured such that Alice and Bob experience losses of 10 dB and 9.2 dB, respectively. (a) The average $S$ value of the CHSH result decreases as the mean photon number $\mu$ increases, entering the classical region $S=2$ starting from $\mu=0.303$. (b) $\Delta S$ showing a decline with increasing $\mu$. (c) The $(S-2)/\Delta S$ as a function of $\mu$ for an interaction time $T_{int}=3$ ns, indicating an optimized state where $(S-2)/\Delta S=91.67$ at $\mu=0.1307$.}
    \label{fig:fig1}
\end{figure}

This allows us to express the value of Eq. \ref{eq:estimate} simply in terms of the $Q_j$. For an angular difference of $\theta$ in the measurement bases, it can be represented as
\begin{eqnarray}\label{eq:EbyQ}
    E(\theta) = \frac{N_{\perp}(\theta)-N_{\parallel}(\theta)}{N_{\perp}(\theta)+N_{\parallel}(\theta)}
    = \frac{2(Q_6-Q_7)}{Q_1-Q_{10}-Q_{11}+Q_{16}}.
\end{eqnarray}
In the case of standard deviation, if the statistical error of the measurement values is given by $\Delta N_{ij}=\alpha\sqrt{N_{ij}}$, then it can be expressed as  
\begin{eqnarray}\label{eq:dEbyQ}
    \Delta E(\theta) &=& \sqrt{\left(\frac{\partial E}{\partial N_{\parallel}}\Delta N_{\parallel}\right)^2+\left(\frac{\partial E}{\partial N_{\perp}}\Delta N_{\perp}\right)^2}\nonumber\\
    &=& \frac{\alpha\sqrt{T_{int}[(Q_{16}-Q_{11}-Q_{10}+Q_1)^2-4(Q_6-Q_7)^2}]}
    {(Q_{16}-Q_{11}-Q_{10}+Q_1)^{3/2}}.
\end{eqnarray}
By substituting Eq. \ref{eq:EbyQ} and Eq. \ref{eq:dEbyQ} into Eq. \ref{eq:chsh} and Eq. \ref{eq:dchsh} according to the appropriate angles, we can finally calculate $\frac{S-2}{\Delta S}$. Using this, Fig. \ref{fig:fig1} shows the trend of $\frac{S-2}{\Delta S}$ according to the mean photon number. These results indicate the existence of an optimized brightness of entangled photon pairs that exhibits stable nonlocality characteristics for the given experimental setup.

\subsection{Experimental setup and results}
\begin{figure}
    \centering
    \includegraphics[width=\linewidth]{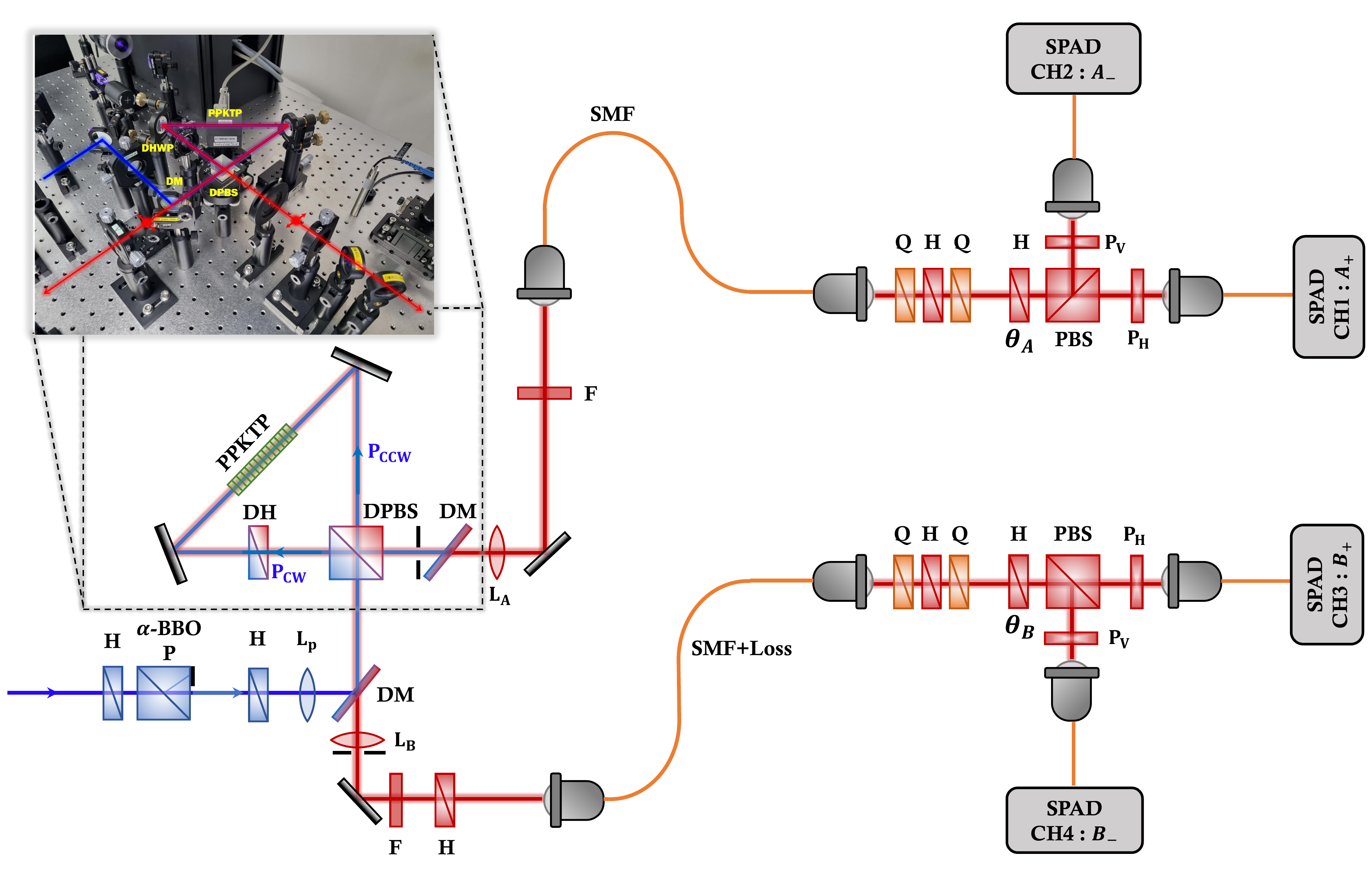}
    \caption{Setup for performing CHSH inequality violation experiments through the distribution of entangled photon pairs. The entangled photon pairs are produced by passing a single-mode laser with a center wavelength of $\lambda_p=403.6$ nm through a half-wave plate (HWP) and an $\alpha$-BBO Glan polarizer, which serves as an attenuator to adjust the intensity of the pump beam. After passing through the attenuator, the pump beam is directed through an HWP rotated at $22.5^\circ$, resulting in diagonal polarization before entering the Sagnac interferometer. Since the interferometer operates at two different wavelengths, careful selection of components for the dual-wavelength HWP (DHWP) and dual-wavelength PBS (DPBS) was undertaken to ensure they function equivalently at both 405 nm and 810 nm. It was verified that the pump beam, with $P = P_{CW} = P_{CCW}$, enters the Type-II PPKTP crystal in a balanced manner. The PPKTP crystal was set to achieve phase matching conditions for collinear type-II SPDC, taking into account the laser pump beam’s central wavelength, with a poling period of 9.75$\mu$m and set at a temperature of $45\pm 0.2^\circ C$ \cite{fan1987second, konig2004extended, wiechmann1993refractive, emanueli2003temperature}. The generated entangled photon pairs exhibit the $\psi^-$ Bell state. However, to facilitate the experiment, an HWP at $45^\circ$ was used to convert it to the $\phi^+$ Bell state. Subsequently, the entangled photon pairs are coupled into a single-mode fiber (SMF) and transmitted to independent measurement units, with the Bob side configured to allow for adjustable losses. After passing through the measurement unit, the entangled photon pairs are compensated with two QWPs and one HWP \cite{simon2012hamilton}. The HWP in front of the PBS allows for the selection of measurement bases by adjusting its angle. Each side, Alice and Bob, is equipped with two single-photon avalanche detectors (SPADs) to simultaneously verify single counts, 2-fold coincidences, and 3-fold coincidences through the nuclear instrumentation module (NIM) and data acquisition (DAQ) systems. The additional components are summarized as follows: H: HWP; $L_p$: Pump lens with a focal length of 200 mm; DPBS: Dual-wavelength PBS; DH: Dual-wavelength HWP; DM: Dichroic mirror; $L_A$ and $L_B$: Lenses set at 250 mm for the Alice and Bob sides; F: 10nm bandpass filter; $P_H$ and $P_V$: Polarizers for horizontal and vertical polarization.
    }
    \label{fig:setup}
\end{figure}
The experimental setup is divided into a Sagnac interferometer for generating polarization-based entangled photon pairs and a measurement part for detecting the distributed entangled photon pairs, as shown in Fig. \ref{fig:setup}. The Sagnac interferometer uses a Type-II PPKTP crystal to produce entangled photon pairs through SPDC at a wavelength of 807.2 nm. A continuous wave laser (TopMode, Toptica) with a wavelength of 403.6 nm and a coherence time of 83 ns was employed as the pump beam for SPDC. For the SPDC process, the pump beam laser passes through a half-wave plate (HWP) at an angle of $22.5^\circ$ to match the phase matching condition in both directions of the 10 mm long PPKTP crystal. The power $P$ of the pump beam entering the PPKTP crystal can reach up to 15 mW in both CW and CCW directions; in this experiment, $P$ was adjusted in increments of 2 mW, ranging from 0.1 mW to 14 mW. The entangled photon pairs generated by the Sagnac interferometer are produced in a $\psi^-$ Bell state according to the Hamiltonian in Eq. \ref{eq:hamiltonian}. However, for the convenience of analysis during the experiment, a HWP set to $22.5^\circ$ was placed in Bob's mode to convert the state into the $\phi^+$ Bell state. Subsequently, both the Alice and Bob channels were filtered through 405 nm bandpass filters with a 10 nm bandwidth and coupled into single-mode fibers (SMF). The entangled photon pairs transmitted through the SMF were polarization-compensated using two quarter-wave plates (QWPs) and a HWP, enabling measurement basis selection via the HWP and a polarizing beam splitter (PBS). Alice and Bob each employed two single-photon avalanche diodes (SPADs) for simultaneous measurements, and recording single counts, two-fold coincidences, and three-fold coincidences. The logic gate of the nuclear instrumentation module (NIM) used for measuring coincidences operated with a coincidence window of $T_{int}=3$ ns, which is significantly shorter than the coherence time of the photon pairs, ensuring the ability to measure coincidences in a single temporal mode.

\begin{figure}
    \centering
    \includegraphics[width=\linewidth]{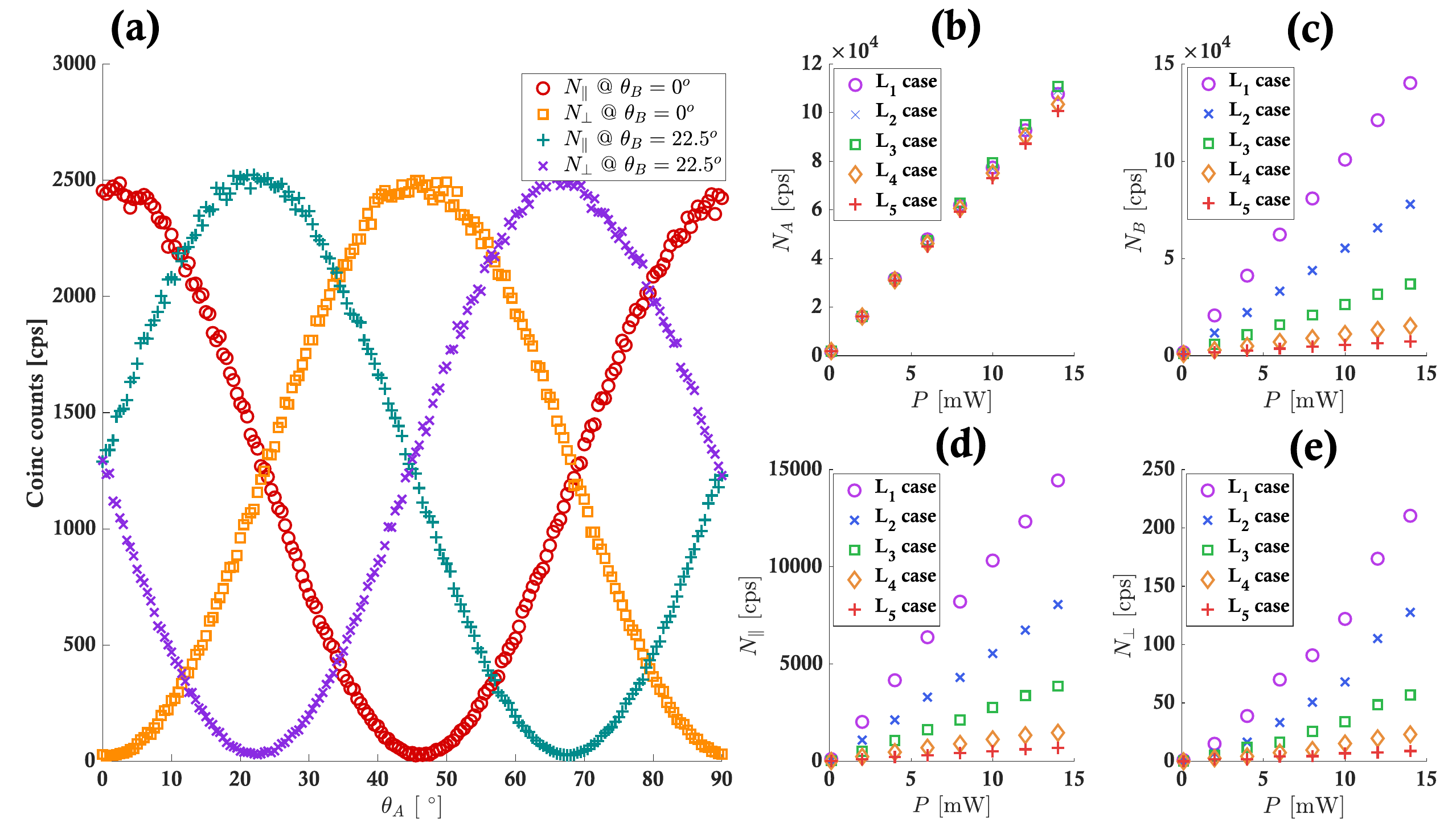}
    \caption{(a) Experiment data showing the conditions for two measurement bases, $\theta_B=0^\circ$ and $\theta_B=22.5^\circ$, along with the visibility of $N_\parallel$ and $N_\perp$ for the entangled photon pairs generated in the Sagnac interferometer. Under the conditions of $P=1.13$ mW, the coincidence counts are shown for $\theta_A$ in $2^\circ$ increments over 1 second, demonstrating a balanced visibility of 0.985 in four cases. (b) Alice's single count $N_A$. (c) Bob's single count $N_B$. (d) Counts of $N_{\parallel}$. (e) Counts of $N_{\perp}$; (b)-(e) were constructed under loss environments of $L_1=-9.03$ dB, $L_2=-11.67$ dB, $L_3=-15.00$ dB, $L_4=-18.39$ dB, and $L_5=-21.74$ dB, with experiments conducted for seven power levels ranging from 0.01 mW to 14 mW in 2 mW increments. Data were collected under the condition of $\theta_A=\theta_B=0^\circ$.}
    \label{fig:fig3}
\end{figure}

For the $\phi^+$ entangled photon pair source, it is necessary that Alice and Bob show the same polarization not only in the rectilinear basis but also in the diagonal basis. This requires careful balancing of the power of clockwise (CW) and counterclockwise (CCW) modes in the Sagnac interferometer, precise phase alignment between Alice and Bob, and balanced efficiency across each SPADs. This balance can be verified by measuring coincidences as Bob rotates his angle by $90^\circ$, while Alice performs measurements in both the rectilinear and diagonal bases, allowing observation of visibilities in the coincidence counts. We demonstrate a visibility of approximately 0.985 at $P=1.13$ mW, as shown in Fig. \ref{fig:fig3}, indicating good balance in each case. Factors influencing the visibility include dark counts, afterpulse, and minor multi-photon effects.

Considering the multi-photon effects, the 16 possible cases for measuring the distribution of entangled photon pairs can be derived as a linear combination according to Eq. \ref{eq:Q-func}. The values of $Q_j$ are influenced by the parameters $\mu$, $\tau_1$, and $\tau_2$ of the experimental setup, making it essential to pre-compute these values for accurate analysis of single and coincidence counts. Analyzing the data from Fig. \ref{fig:fig3}(e) indicated that at $P=0.1\sim2$ mW, the low contribution from $N_{\perp}$ suggests that multi-photon events are negligible. In this scenario, we can estimate $\mu$ from the single counts and coincidence counts, as we have $N_\parallel = \mu\tau_A\tau_B/T_{int}$, $N_A = \mu\tau_A/T_{int}$, and $N_B = \mu\tau_B/T_{int}$. With an average single count of 20,717 cps for Alice and 16,135 cps for Bob, the average coincidence count $N_\parallel$ is measured at 2,024 cps, resulting in $\tau_A = 0.10$ and $\tau_B = 0.13$. For Bob's mode, experiments were conducted under five different loss conditions by additionally increasing the loss by approximately 3 dB. The mean photon number during the time $T_{int}$ follows
\begin{equation}\label{eq:mu}
    \frac{\mu}{T_{int}}=\left(\frac{N_A}{2\tau_A}+\frac{N_B}{2\tau_B}\right),
\end{equation}
which allows us to obtain $\mu$ for low $P$ where multi-photon events are nearly negligible. Since $\gamma = C_\gamma\sqrt{P}$, we can also estimate the mean photon number at higher powers, where the multi-photon effect can no longer be ignored, using the new equation $\mu = \sinh^2(C_\gamma\sqrt{P})$. In our experimental setup, $C_\gamma$ was determined to be 0.0469, enabling us to estimate the $\mu$ values for each $P$ controlled through the attenuator.

\begin{figure}
    \centering
    \includegraphics[width=\linewidth]{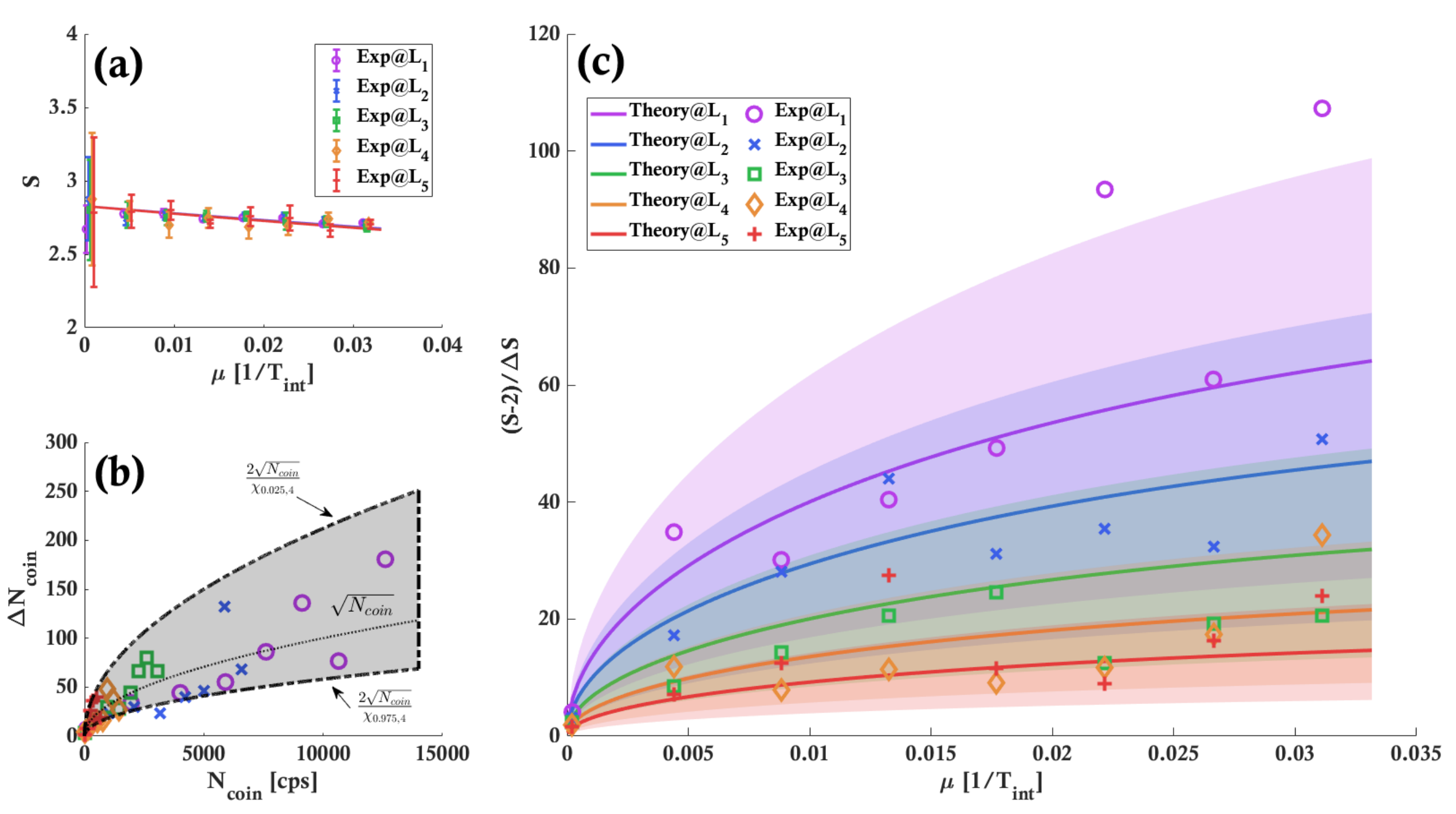}
    \caption{(a) The average and standard deviation of the CHSH values obtained from seven pump beam intensities and five loss environments are presented, confirming consistency with Fig. \ref{fig:fig1}(a) and (b). (b) Each case demonstrates that the coincidence count follows the form $\Delta N_{coinc} = \alpha \sqrt{N_{coin}}$. The data is distributed within approximately a 90$\%$ confidence interval. (c) The experimental values for $(S-2)/\Delta S$ generally align with Fig. \ref{fig:fig1}(c); however, it exhibits a statistical property where the $\alpha$ values are not constant but rather fall within a range.}
    \label{fig:fig4}
\end{figure}

The CHSH value, as indicated in Eq. \ref{eq:CHSH}, requires the calculation of $E(\theta)$ for four different conditions according to Eq. \ref{eq:estimate}. However, the angles mentioned for Alice and Bob refer to the measurement bases of mathematical states, assuming an experiment that utilizes polarizers. In our setup, where HWPs and PBSs are employed to select measurement bases, the rotation angles of the HWP must be set to $\varphi'{A1}=0^\circ$, $\varphi'{A2}=22.5^\circ$, $\varphi'{B1}=11.25^\circ$, and $\varphi'{B2}=33.75^\circ$. To investigate how the averages and standard deviations of the CHSH values differ across various environments, experiments were conducted under five loss conditions and seven pump beam intensities, with the results shown in Fig. \ref{fig:fig4}(a). The CHSH value $S$ starts at approximately 2.8 for low pump beam intensity and decreases to 2.65 at 14 mW. While higher losses resulted in reduced CHSH values compared to lower losses, the overall changes were not significant. Similarly, as the pump beam intensity increases, the standard deviation, as predicted in Fig. \ref{fig:fig1}(b), decreases, although the differences due to losses are minimal. 

The value of $(S-2)/\Delta S$ not only indicates how far the source deviates from classical behavior but also reflects the stability of the entanglement of the distributed photon pairs influenced by environmental characteristics. As the loss increases, $(S-2)/\Delta S$ decreases, as seen in Fig. \ref{fig:fig4}(c). Unlike merely observing $S$ or $\Delta S$, this metric provides a clearer understanding of the differences across various environments. Interestingly, instead of following Poisson statistics, the statistical errors of the measurements exhibited a 90$\%$ confidence interval range, as depicted in Fig. \ref{fig:fig4}(b). This characteristic also affects the value of $(S-2)/\Delta S$. As observed in Fig. \ref{fig:fig4}(c), the experimental results match well with the predicted $(S-2)/\Delta S$ obtained by substituting $\alpha$ values from Eq. \ref{eq:dEbyQ} with $2/\chi_{0.975,4}$ and $2/\chi_{0.025,4}$, demonstrating that our proposed mathematical modeling accurately describes the experimental results. Especially, in an environment where the total loss of the entanglement distribution setup is $-19.03$ dB and $\mu$ is 0.026, the value of $S$ decreased to 2.69, while $(S-2)/\Delta S$ rose to 60.95.

\section{Conclusion}
In conclusion, we have demonstrated that our previous theoretical research findings\cite{kim2024strategy}, along with the predictions related to the violation of the CHSH inequality, align well with the experimental results. Although the violation of the CHSH inequality in experiments can be derived through relationships if visibility can be measured, the visibility itself is significantly influenced by various experimental factors. Therefore, predicting the results of CHSH inequality violations based on the experimental setup prior to conducting the actual experiment is not straightforward. This study focused on the multi-photon effect among several factors affecting visibility, providing a mathematical modeling of its impact on experimental outcomes. Notably, since changes in the CHSH value do not exhibit significant differences across various loss environments, we proposed a theoretical model to calculate the changes in $(S-2)/\Delta S$ along with experimental evidence. In cases where the intensity of entangled photon pairs is low, the distributed photon pairs can exhibit a high $S$ value, but with a high $\Delta S$, leading to the possibility of $S > 2\sqrt{2}$, which reduces reliability. Conversely, when the intensity of the entangled photon pairs is high, $\Delta S$ decreases, resulting in increased reliability for CHSH experiments; however, the $S$ value itself decreases due to errors arising from the multi-photon effect. As such, the trade-off between $S$ and $\Delta S$ with respect to the intensity of the entangled photon pairs suggests that there is an optimal brightness that can maximize $(S-2)/\Delta S$. Unfortunately, our experiment was limited by the constraints of nonlinear gain, preventing us from confirming this expectation. Theoretically, it is anticipated that the mean photon number should be 0.13 when the losses for Alice and Bob are both 10 dB. Achieving this requires significantly higher nonlinear gain, which necessitates either a substantial increase in laser power or the use of materials with a higher nonlinear gain than Type-II PPKTP. Recent studies on ultrabright entangled photon pair sources indicate that Type-0 PPKTP exhibits more than 80 times greater nonlinear effects than Type-II PPKTP \cite{steinlechner2014efficient, cao2018bell, park2024ultrabright}, suggesting that practical applications could indeed be feasible.

The mathematical modeling and experimental evidence proposed in this paper are expected to be beneficial not only for Bell inequality violation experiments utilizing high-loss regime and high-brightness entangled photon pair sources but also for the design of various practical applications \cite{bennett1993teleporting, gottesman2012longer, kim2024purification}. Since an optimized brightness exists for various loss environments, it enables the prior preparation of the required brightness of entangled photon pairs for designing quantum repeater experiments over given distances. Furthermore, if entangled photon pair sources are already prepared, this study could be appropriately used to assess the maximum distances between nodes when building a quantum network, aiding in cost management and other considerations.

\begin{acknowledgments}
This research was supported by the Challengeable Future Defense Technology Research and Development Program through the Agency For Defense Development (ADD) funded by the Defense Acquisition Program Administration (DAPA) in 2023 (No.915027201). 
\end{acknowledgments}


\bibliography{ldmpref}

\end{document}